\documentclass[conference]{IEEEtran}
\usepackage{graphicx,array,amsmath,caption,epstopdf,amsfonts,algorithm,algorithmic,hyperref}
\usepackage{subfigure,psfrag,epsfig,amssymb,array,flushend}
\hypersetup{pdfborder={0 0 0},colorlinks=false, linkcolor=black, filecolor=black, citecolor=black}
\begin{document}
\title{A blind Robust Image Watermarking Approach
exploiting the DFT Magnitude}
\author{
    \IEEEauthorblockN{Mohamed HAMIDI\IEEEauthorrefmark{1}, Mohamed EL HAZITI
\IEEEauthorrefmark{4}, Hocine CHERIFI    
    \IEEEauthorrefmark{7},Driss ABOUTAJDINE\IEEEauthorrefmark{1}}
    \IEEEauthorblockA{\IEEEauthorrefmark{1} Associated Unit to the CNRST-URAC N 29,\\ Faculty of Sciences, University of Mohammed V,\\BP 1014 Rabat, Morocco
    \\{hamidi.medinfo@gmail.com } , {aboutaj@ieee.org  }
    \IEEEauthorblockA{\IEEEauthorrefmark{4} Higher School of Technology, Sale, Morocco
    \\{elhazitim@gmail.com } 
    \IEEEauthorblockA{\IEEEauthorrefmark{7}Laboratoire Electronique, Informatique et Image (Le2i) UMR 6306 CNRS, \\University of Burgundy, Dijon, France    \\{hocine.cherifi@u-bourgogne.fr }  
}}}}
\maketitle
\begin{abstract}
Due to the current progress in Internet, digital contents (video, audio and images) are widely used. Distribution of multimedia contents is now faster and it allows for easy unauthorized reproduction of information. Digital watermarking came up while trying to solve this problem. Its main idea is to embed a watermark into a host digital content without affecting its quality. Moreover, watermarking can be used in several applications such as authentication, copy control, indexation, Copyright protection, etc. \\
In this paper, we propose a blind robust image watermarking approach as a solution to the problem of copyright protection of digital images. The underlying concept of our method is to apply a discrete cosine transform (DCT) to the magnitude resulting from a discrete Fourier transform (DFT) applied to the original image. Then, the watermark is embedded by modifying the coefficients of the DCT using a secret key to increase security. 
Experimental results show the robustness of the proposed technique to a wide range of common attacks, e.g., Low-Pass Gaussian Filtering, JPEG compression, Gaussian noise, salt \& pepper noise, Gaussian Smoothing and Histogram equalization. The proposed method achieves a Peak signal-to-noise-ration (PSNR) value greater than 66 (dB) and ensures a perfect watermark extraction.
\end{abstract}
\begin{IEEEkeywords}
copyright protection, image watermarking, Discrete Fourier Transform (DFT), Discrete Cosine Transform (DCT), embedded, blind, etc.
\end{IEEEkeywords}
\IEEEpeerreviewmaketitle
\section{Introduction}
\IEEEPARstart{D}{ue} to the huge growth of electronic publishing industry, multimedia data can be distributed and copied much  easier. The recent rapid development of new technologies for multimedia services, has resulted in a strong demand for reliable and secure protection techniques for multimedia data. Digital image watermarking, especially robust image watermarking, came up while trying to solve this problem. The main idea behind image watermarking is to embed a watermark within multimedia contents (image, video, audio, etc). The watermark must be imperceptible, so that it does not affect the quality of the host data and it should be difficult or even impossible for unauthorized users to counterfeit or remove it. 

The process of watermarking is carried out with three key requirements at hand \cite{cox2007digital}  : the imperceptibility, the robustness and the capacity. A good  watermarking system must provide the best tradeoff between these three aspects.
 
The most important requirement of an image watermarking system is imperceptibility. Indeed, if this system fails to achieve this requirement it will not be suitable for practical applications.	
Imperceptibility refers to perceptual similarity between the host image before watermarking process and the watermarked image. Therefore, a good watermarking scheme causes no artifacts or quality loss in the images. Whereas, the robustness is the ability of detecting the watermark even if the original image has incurred changes such as filtering, lossy compression, scaling, rotation, etc. 
Besides, the capacity requirement refers to the maximum number of bits that can be embedded in a given host data.  
 Based on the resistance to attacks, digital image watermarking algorithms can be divided into fragile \cite{xiao2012improved}, semi-fragile \cite{qi2011quantization} and robust watermarking \cite{zhang2012improved}.
The existing algorithms can be also distinguished according to the domain the watermark is embedded in. There are two main domains : spatial domain  and frequency domain. Techniques operating in the spatial domain \cite{728421} embed the watermark by directly modifying the gray level of image pixels, whereas in frequency domain \cite{650120} a transformation is first applied to the original image and then embedding the watermark into DCT \cite{6116241}, DWT \cite{5560822} or DFT coefficients \cite{6811181}\cite{6256390} .\par
In this paper a  blind watermarking algorithm for digital images is presented. The method, which operates in the frequency domain, embeds the watermark bits in a selected set of discrete cosine transform (DCT) coefficients of the magnitude after carrying out the discrete Fourier transform (DFT) of the host image. The proposed method is designed to be robust against several attacks such as : Low-Pass Gaussian Filtering, JPEG compression, Gaussian noise, salt \& pepper noise, Gaussian Smoothing, etc. Furthermore, we choose to compare our method with the  recent schemes presented in \cite{mingzhi2013combined} and \cite{behloul2014blind}  in order to evaluate it in terms of imperceptibility and robustness.\par
 This paper is organized as follows. Section II discusses the related works. Section III develops the proposed watermarking scheme. Section IV shows the experimental results and section V concludes the paper.

\section{Previous Works}

Recent watermarking studies turned their attention to the frequency domain techniques rather than the spatial domain techniques. Effectively, the transform domain approaches  are more robust compared to  spatial domain  approaches \cite{mingzhi2013combined}\cite{behloul2014blind}\cite{jia2014novel}. Therefore, they are mostly used for robust watermarking \cite{1560462} .\par
In \cite{zhang2012improved}, an improved watermarking algorithm based on DCT is proposed. Firstly, to increase security, the watermark is extended to a size four times larger than the initial watermark. Second,  the watermark is encrypted by using the sine chaotic system . Zhang \textit{et al.} proposed an  adaptive embedding method which embedded the watermark into selected DCT coefficients. In fact, for more robustness, the smaller coefficients after attacks were selected for embedding the watermark. \par
In \cite{1699615}, a digital image watermarking scheme using fast Hadamard transform (FHT) and singular value decomposition (SVD) is presented. Abdallah \textit{et al.}, divide first the host image into small blocks, then they carry out the FHT to each block and  distribute the singular values of the watermark image over the transformed blocks. The singular values of the watermark image are embedded in the DC components of the FHT blocks of the original image. \par 

The same authors proposed in \cite{2007JEI....16c3020A} a robust image watermarking method by carrying out the fast Hadamard transformation (FHT) to small blocks computed from the four discrete wavelet transform (DWT) subbands. The embedding process consists of four main steps. First,  the original image is decomposed into four subbands. After, the four subbands are divided into blocks. Then, the FHT is applied to each block. Finally, the singular value decomposition (SVD) is applied to the watermark image prior to distributing the singular values over the DC components of the transformed blocks.

\par
From another aspect, the blind watermarking techniques receives more attention.
In \cite{jia2014novel}, Shao-li Jia \textit{et al.} proposed a blind watermarking method based on singular value decomposition (SVD). In fact, the SVD is applied to each $4 \times 4$ block of each color component (R,G and B) of the host image.
The embedding scheme consists to modify the second and the third row element of the first column of the orthogonal matrix U according to the watermark information.
 To extract the watermark, the modified relation is utilized. \par
In \cite{mingzhi2013combined}, Mingzhi \textit{et al.} proposed a robust watermarking combining  DWT and DCT. In the embedding process, after decomposing the cover image by a 2-level DWT,  the HL2 sub-band coefficient was divided into 4x4 blocks. Then, for each block, the DCT was performed. The embedding is performed by inserting  two pseudo-random sequences, according to the watermark bit, on the middle band coefficients of DCT. The extracting process is similar to the embedding scheme. In fact, watermarked image was decomposed with 2-level DWT and DCT, then correlation between middle band coefficients of block DCT and the two pseudo-random sequences was calculated to decide which bit was embedded 0 or 1.

Later,  Behloul \textit{et al.} \cite{behloul2014blind} proposed a  blind digital image watermarking technique based on Integer Wavelet Transform (IWT)  and state coding. The approach consists to insert a grayscale image into a color image. Firstly, the  Speeded Up Robust Features (SURF) detector is utilized to extract interest points. Secondly, the IWT transformation is applied  and the watermark is embedded into the local non overlapping areas around each interest point.  The extracting scheme is blind, neither the original image nor the watermark are needed,  only the SURF detector and the state coding are needed to extract the embedded watermark.

In \cite{6811181}, Urvoy \textit{et al.} proposed a blind and robust DFT watermarking scheme based on Human Visual System (HVS) .
The approach, which operates in the Fourier domain, embed the watermark by substitution of both the magnitude (energy) and the phase (information). The watermark strength is adjusted by the amplitude component while the phase component holds the information. With the aim of determining the optimal strength at which the watermark reaches the visibility threshold, a perceptual model of the (HVS) based on local contrast pooling and  Contrast Sensitivity Function (CSF) is used. The method is robust to various kind of attacks, especially geometrical distortions. 

\section{Proposed scheme}
 Our method is inspired by the experiment proposed by Oppenheim \textit{et al.} that shows the importance of the phase compared to the magnitude \cite{1456290}.
 They proved that the information carried by the phase of the image appears to be much more significant than the magnitude. For this reason, we choose to insert the watermark in the magnitude information as it does not influence the image quality.  
 
\subsection{Watermark Embedding}
\begin{figure}[!h]
\captionsetup{justification=centering}
\begin{center}
\includegraphics[scale=0.55]{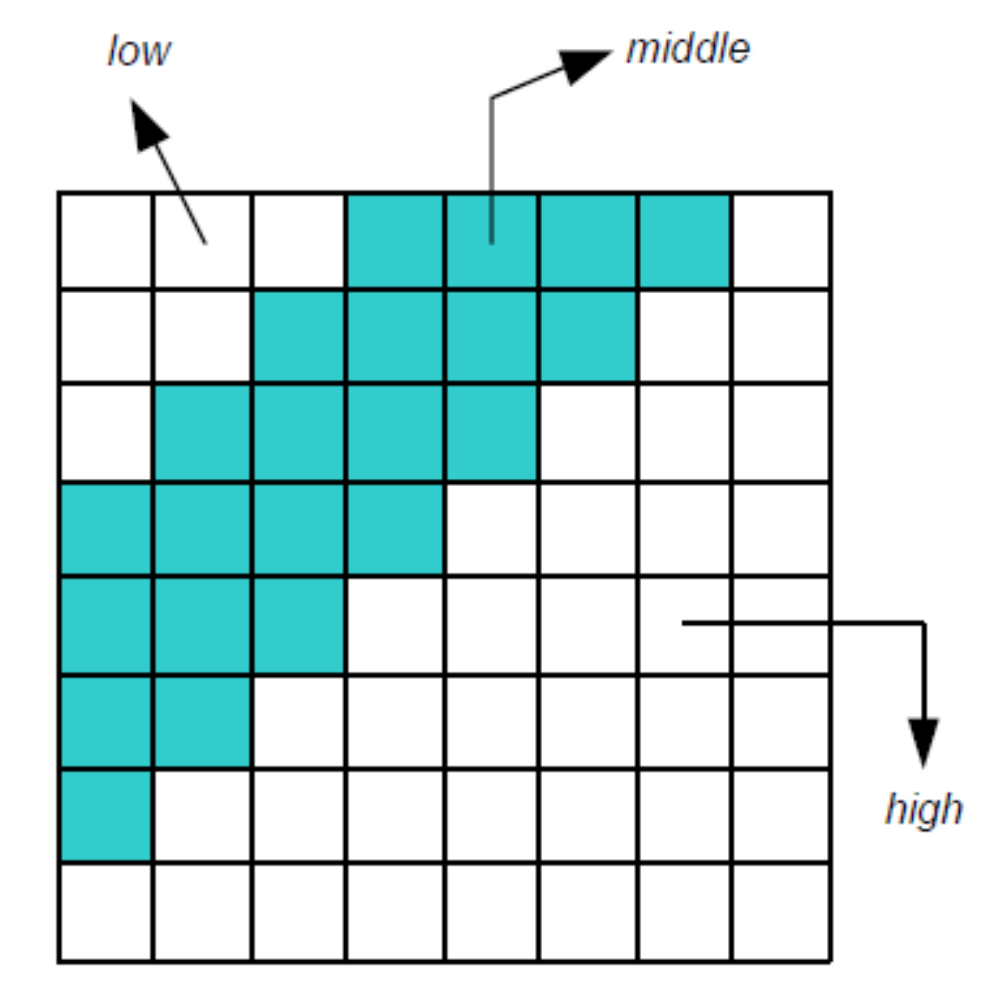}
\end{center}
\caption{The middle band DCT coefficients where the watermark is embedded.}
\label{fig:middlebandfrequencies.}
\end{figure}

Let $F(u,v)$ the Discrete Fourier Transform of an image. The Fourier magnitude and phase are represented as follows : 
\begin{equation}
M(u,v)= \left | F(u,v) \right | = [R^{2}(x,y)+I^2(x,y)]^{1/2}
\end{equation}
\begin{equation}
\phi (u,v)= tan^{-1}\left [ \frac{I(u,v)}{R(u,v)} \right ]
\end{equation}
Where $R(u,v)$ and $I(u,v)$ are respectively the real and imaginary parts of $F(u,v)$.\\

Firstly, the DFT is applied to the original image then the the magnitude $M(u,v)$ and the phase $\phi (u,v)$   are calculated as shown in (1) and (2). Secondly, the magnitude matrix is divided into square blocks of size $8 \times 8$. Then, we apply the DCT on every block of the  magnitude. Next, two uncorrelated pseudo-random sequences using a secret key are generated : one sequence for  "0" bits (PN\_Seq\_0) and another  sequence for the "1" bits (PN\_Seq\_1).  Note that each  PN sequence must have the same number of elements that the number of middle band coefficients. \par
 The embedding process consists of embedding PN-Sequences depending on the bit of the watermark. The middle band coefficients (see the colored region in Fig. \ref{fig:middlebandfrequencies.}) of the DCT transform of the DFT magnitude are commonly used for watermark embedding to avoid modifiying the important visual parts of image.\par
  The watermark strength is handled by the gain factor $k$ which controls the tradeoff between robustness and imperceptibility. In fact, an increase of the gain factor increases the watermarking robustness while it decreases the imperceptibility of the watermark. Thus, we take a value of $k$ so that we have a good tradeoff between robustness and imperceptibility. Then, inverse DCT is applied to obtain the modified  magnitude. Finally, the watermarked image is reconstructed with the unchanged phase and the modified magnitude using equation (4). Afterwards, the inverse discrete Fourier transform (IDFT) is performed to obtain the watermarked image pixel values.\\
Fig. 2 sketches the watermark embedding process which is described in detail Algorithm \ref{alg:embedding-process}.  

\begin{algorithm}[t]
\caption{Watermark Embedding}
\label{alg:embedding-process}
\begin{algorithmic} 
\REQUIRE Originale image, Watermark, Key, PN\_Seq\_0, PN\_Seq\_1.
\ENSURE Watermarked image.
\STATE 1. Apply DFT to the original image and calculate the magnitude and the phase.
\STATE 2. Generate two uncorrelated PN sequences for middle frequency band coefficients using a secrete key. 
\STATE 3. Divide the magnitude of DFT into $8 \times 8$ blocs and then apply the DCT.
\STATE 4. Insert the two PN sequences bits according  to watermark bits using the equation (3): \\
 \textbf{if} Watermark (bit) $= 0$ \hspace*{0.1cm}  \textbf{then}  \hspace*{0.1cm} $W(u,v)=$ PN\_Seq\_0\\ \hspace*{0.2cm} \textbf{else}  \hspace*{0.5cm} $W(u,v)=$ PN\_Seq\_1.
\STATE 5. Perform IDCT on each watermarked magnitude block.
\STATE 6. Reconstruct the watermarked image with the modified magnitude using the equation (4) \cite{1456290}. 
\STATE  7. The final watermarked image is obtained  by performing the  IDFT.
\end{algorithmic}
\end{algorithm}
\begin{equation}
M_w(u,v)=\begin{Bmatrix}
M(u,v)+k*W(u,v) \hspace*{0.5cm} u,v \in F_M \\ 
M(u,v) \hspace*{2.5cm} u,v \notin F_M         
\end{Bmatrix}
\end{equation}
 Where  $M_w(u,v) $ is the watermarked magnitude block, $M(u,v)$  represents the $8 \times 8$ DCT block of the magnitude, $F_M$ is the middle frequency band coefficients, $k$ controls the watermark strength (gain factor). 
\begin{equation}
\hspace*{-1cm}
I_w(u,v) = M_w(u,v)*e^{(j\phi (u,v))}
\end{equation}

\begin{figure}[t]
\captionsetup{justification=centering}
\begin{center}
\includegraphics[scale=0.60]{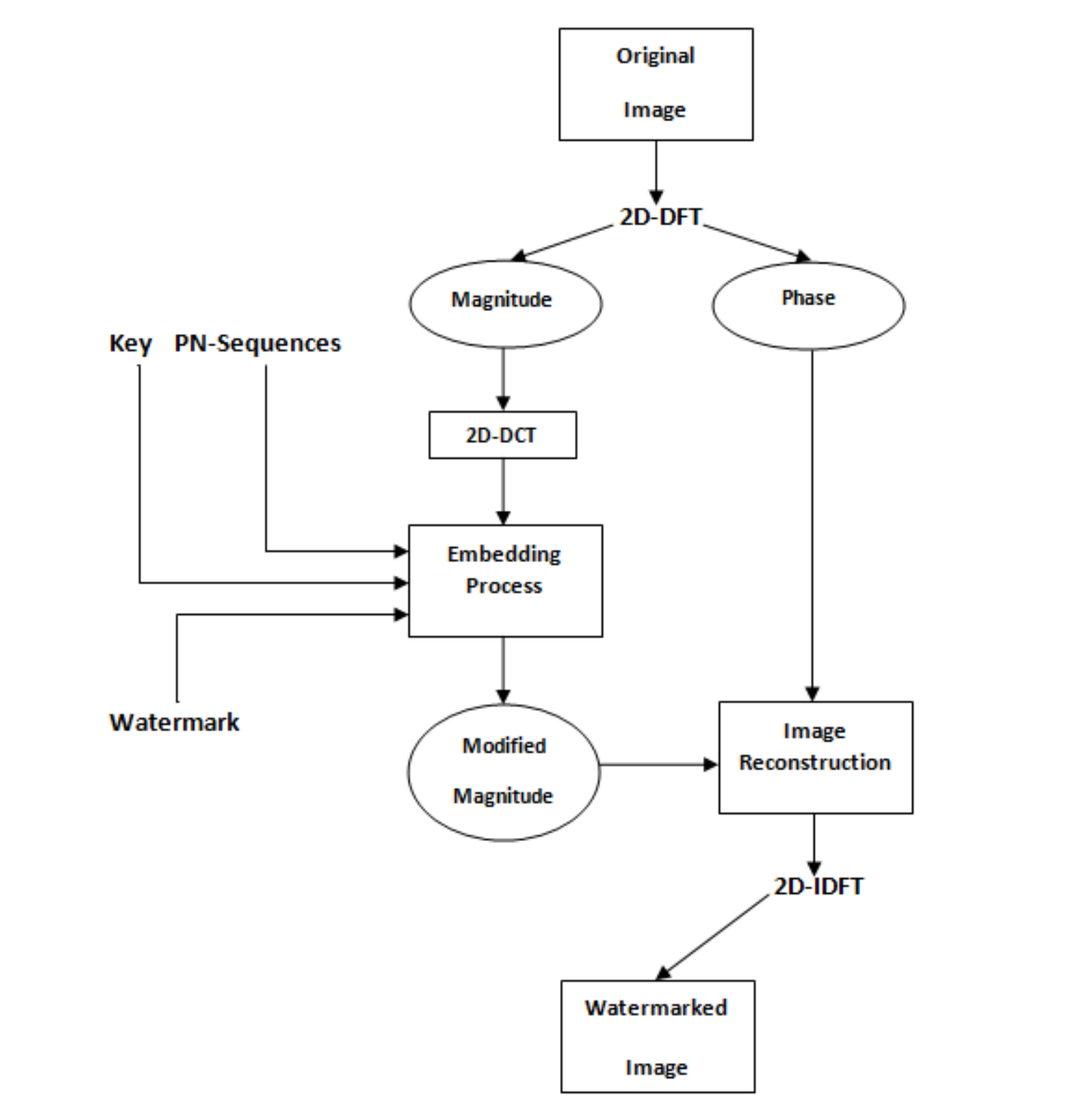}
\end{center}
\caption{ Embedding scheme.}
\end{figure}

\begin{algorithm}[t]
\caption{Watermark Extracting}
\label{alg:WatermarkExtracting}
\begin{algorithmic} 
\REQUIRE Watermarked image.
\ENSURE  Watermark.
\STATE 1. Apply DFT to the watermarked image and calculate the magnitude.
\STATE 2. Generate two PN sequences (PN\_Seq\_0 and PN\_Seq\_1) using the same secrete key used in the embedding process.  
\STATE 3. Apply DCT to the Magnitude of DFT and extract  the  middle  frequency  band coefficients.
\STATE 4. Calculate  the  correlation  between  the middle frequency band coefficients $F_M$ and the two PN sequences.
\STATE 5. Extract the $i^{th}$ Watermark bit $W_i$ as follows : 
\begin{equation}
W_i=\begin{Bmatrix}
0 \hspace*{0.5cm}  if \hspace*{0.5cm}  Corr(0) > Corr(1) \\ 
1 \hspace*{0.5cm}  if \hspace*{0.5cm}  Corr(1) > Corr(0) 
\end{Bmatrix}
\end{equation}
Where $Corr(0)$ is  the  correlation  between the middle frequency band coefficients of  $i^{th}$  block and PN\_Seq\_0, and $Corr(1) $  is  the  correlation  between  the middle frequency band coefficients of  $i^{th}$  block and PN\_Seq\_1.
\end{algorithmic}
\end{algorithm}
\subsection{Watermark Extraction}
\begin{figure}[t]
\centering
\includegraphics[scale=0.75]{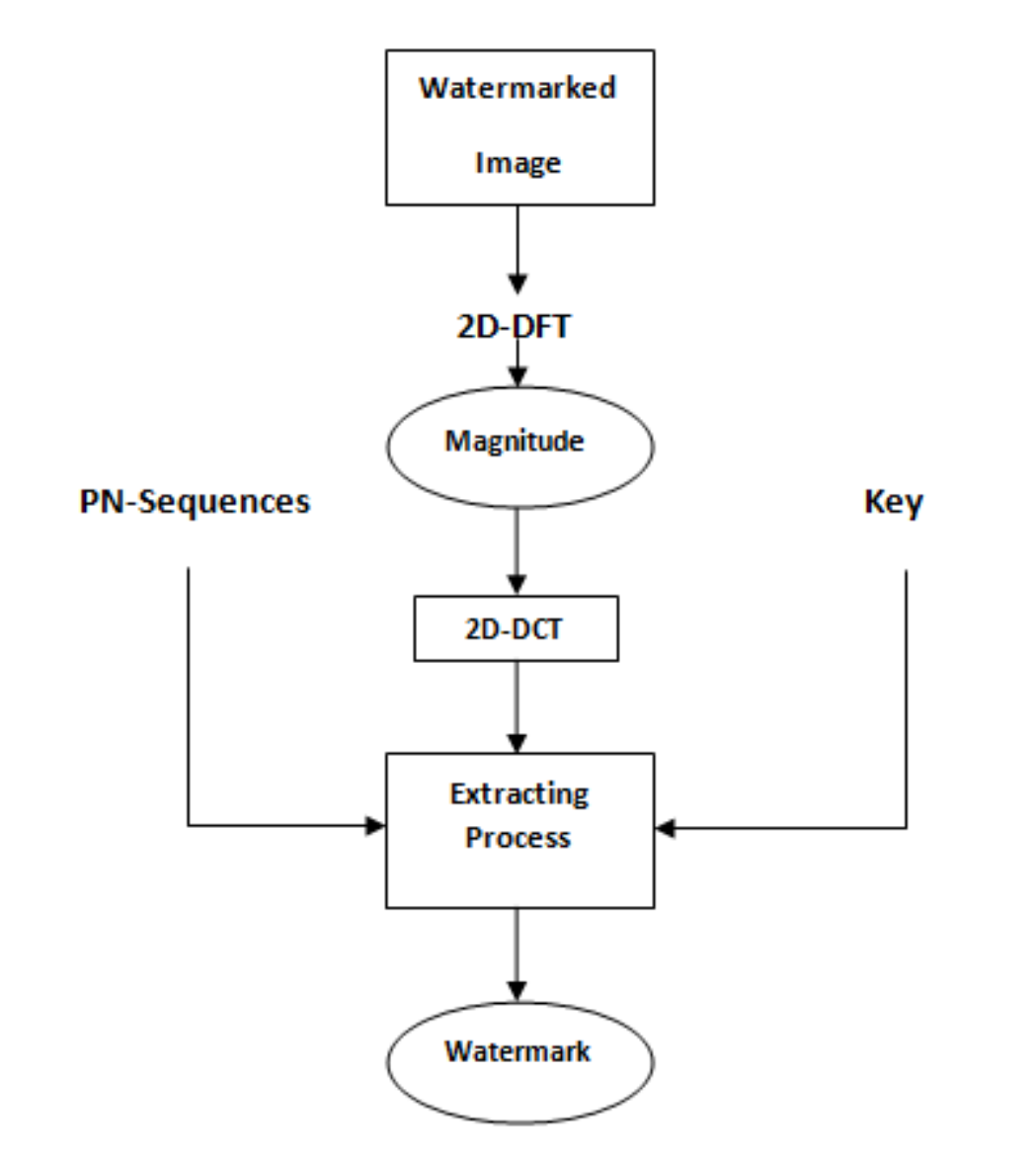}
\caption{Extracting scheme.}
\label{fig:Extracting scheme.}
\end{figure}

As we rely on a blind watermarking scheme, the extraction of the watermark does not need the original image neither the watermark.
Only the secret key is needed for the extraction phase, see Fig. \ref{fig:Extracting scheme.}. \\
After applying the 2D-DFT to the watermarked image and calculating the DFT magnitude, we generate two PN Sequences with the same secret key. As a consequence, we obtain the same PN sequences as those of the embedding scheme. Then, we perform the 2D-DCT to the DFT magnitude. In the extracting scheme, as shown in Fig. \ref{fig:Extracting scheme.}, the middle-band frequencies coefficients of each $8 \times 8$ DCT bloc is extracted. Then, for each bloc, we calculate the correlation between the middle band frequencies coefficients and the two PN sequences. Finally, we extract the $i^{th}$ watermark bit using equation (5).  The proposed extraction scheme is further described in the Algorithm \ref{alg:WatermarkExtracting}.

\section{Experimental Results}
In this section, with the aim of evaluating the performance of the proposed scheme, various experiments  are presented. The gray scale $(512 \times 512)$ "Lena", "Mandril","Goldhill","Peppers" and "Cameraman" images are used as host images, as depicted in Fig. 3. A $(19 \times 52)$  binary logo is used as watermark  as shown in Fig. 4. We choose to compare our scheme  with  schemes presented in \cite{mingzhi2013combined} and \cite{behloul2014blind}, in terms of  imperceptibility and robustness. For this purpose, another $(64 \times 64)$ binary image is used as watermark.
\subsection{Imperceptibility}
Peak Signal to Noise Ratio (PSNR) is used to evaluate the perceptual change between the original image and the watermarked one. It can be defined as follows : \\
\begin{equation}
PSNR= 10 \log (\frac{MAX^2}{MSE})
\end{equation}
Where MAX is the maximum possible pixel value of the image, and MSE is given by : 
\begin{equation}
MSE=\frac{1}{mn} \sum_{i=0}^{m-1}\sum_{j=0}^{n-1} [I(i,j)-K(i,j)]^2
\end{equation}
\begin{figure}
    \centering
    \subfigure[]{\label{sub1} \includegraphics[width=1.56cm]{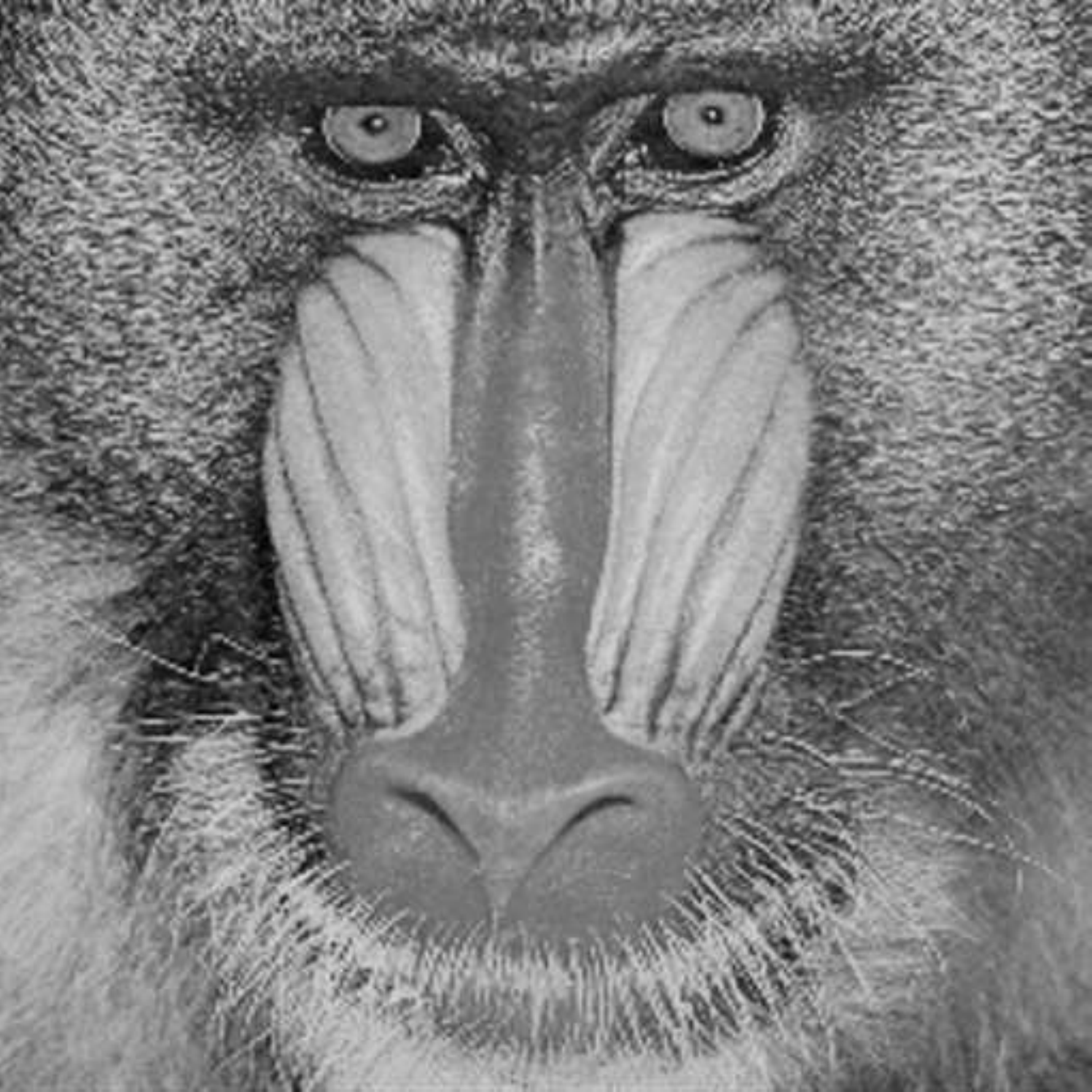}}
    \subfigure[]{\label{sub2} \includegraphics[width=1.56cm]{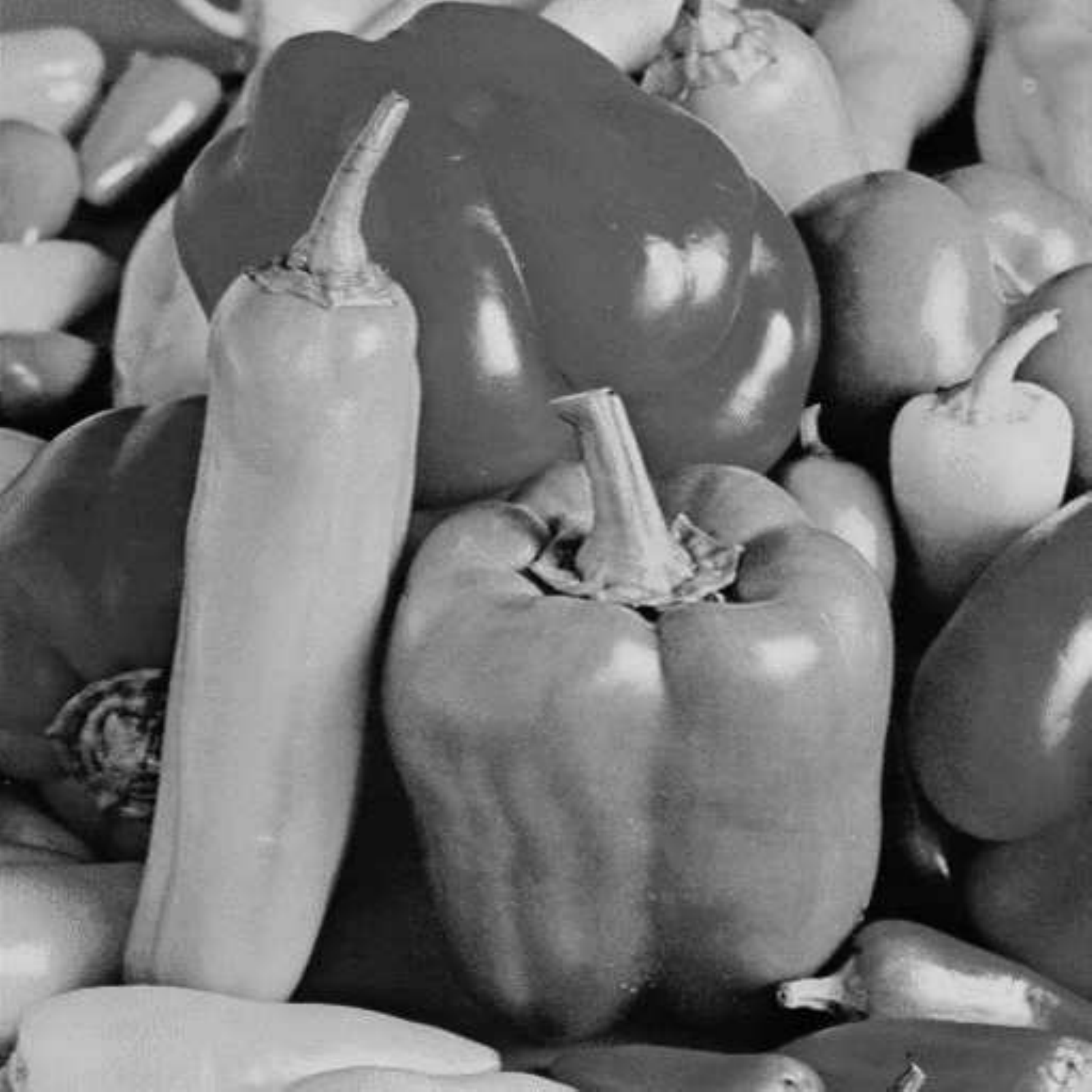}}
    \subfigure[]{\label{sub3} \includegraphics[width=1.56cm]{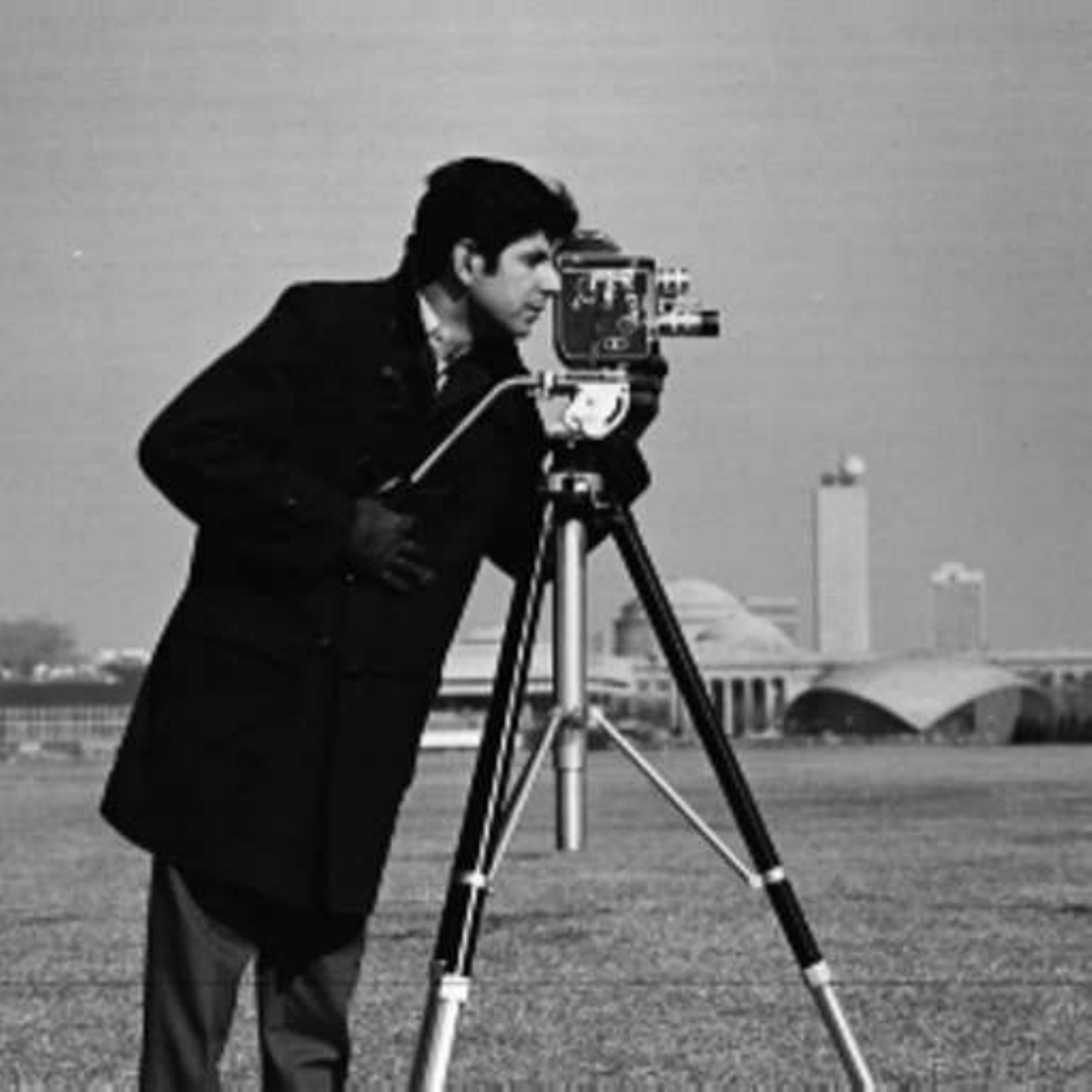}}
    \subfigure[]{\label{sub4} \includegraphics[width=1.56cm]{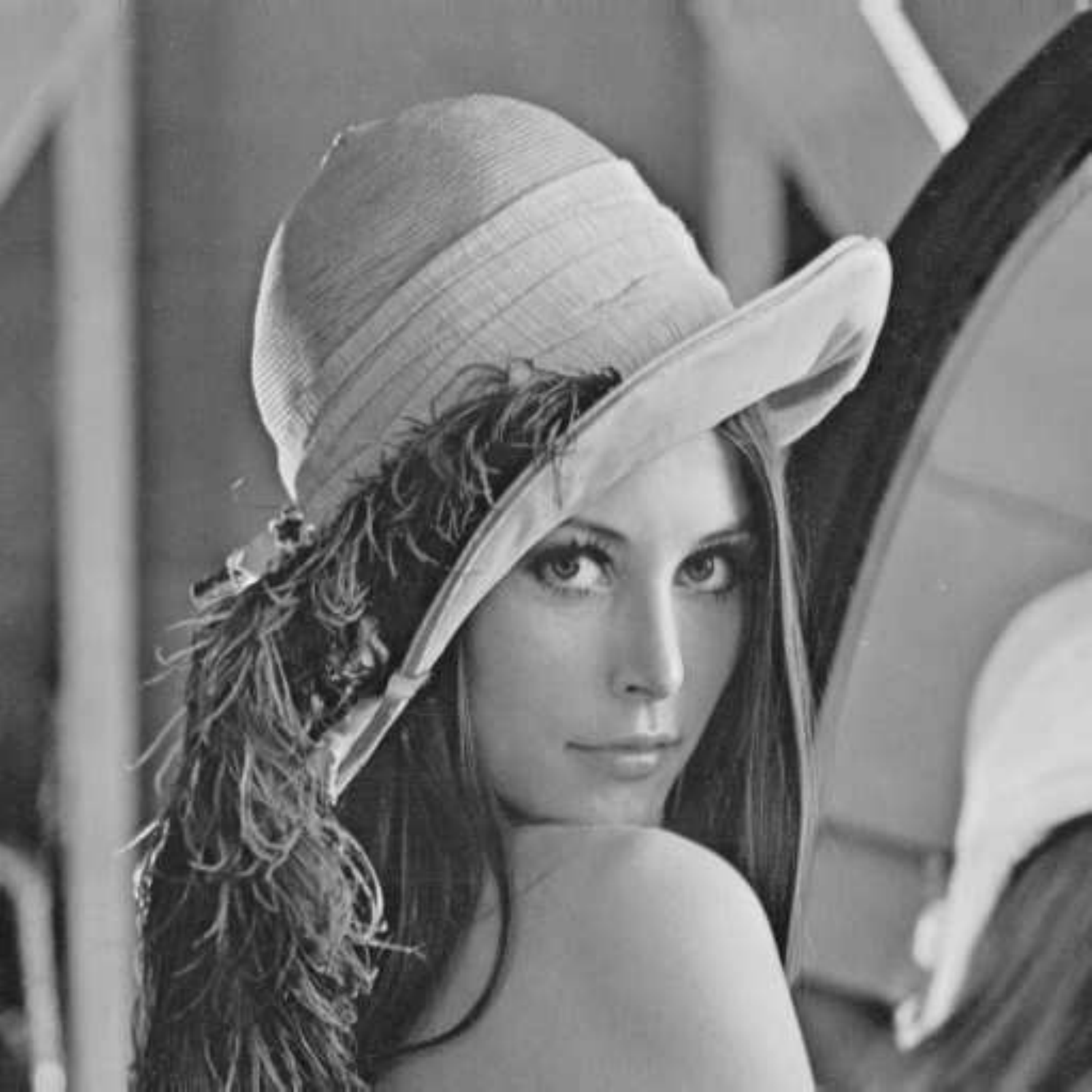}}
\subfigure[]{\label{sub5} \includegraphics[width=1.56cm]{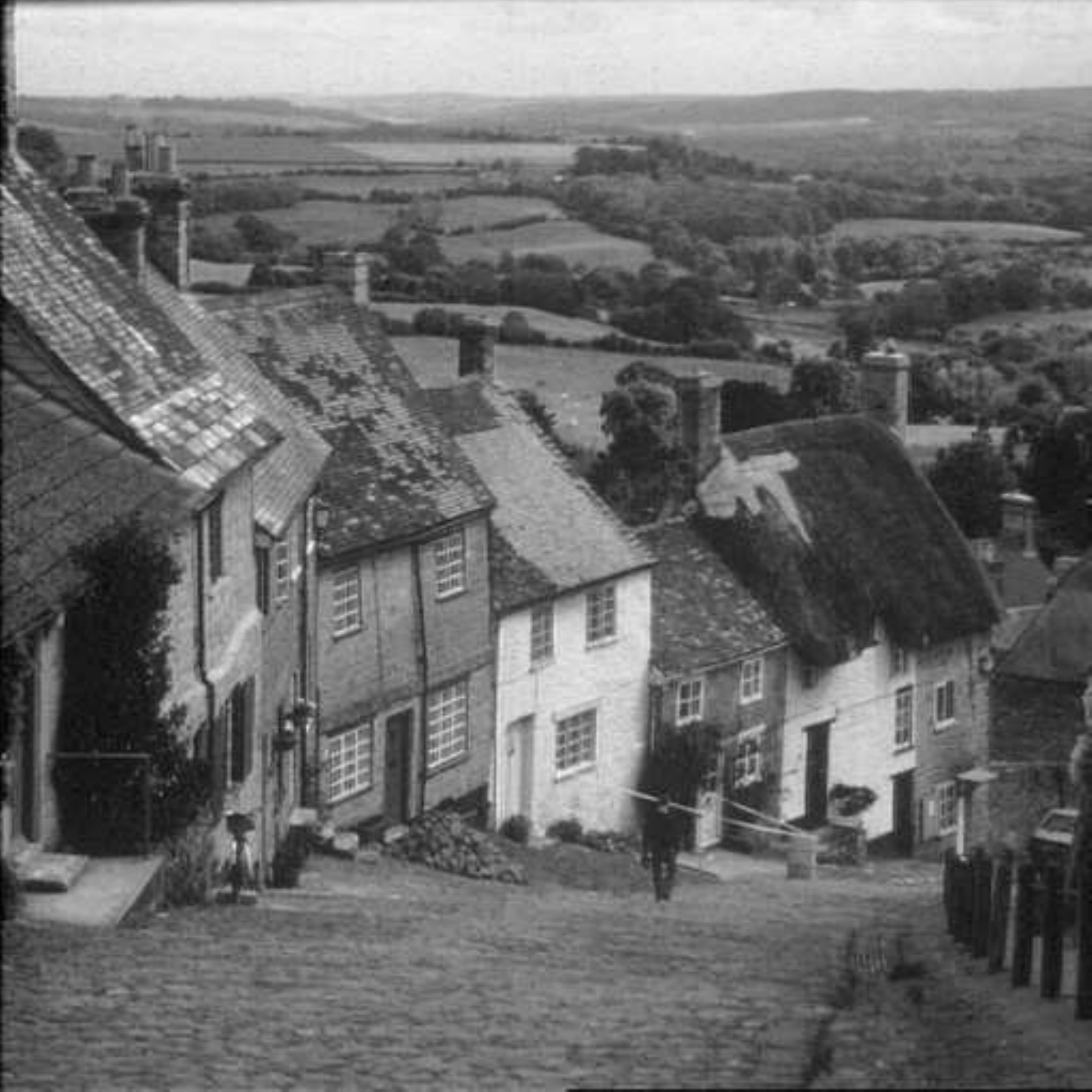}}  

\subfigure[]{\label{sub6} \includegraphics[width=1.56cm]{figs/mandril_gray.pdf}}
    \subfigure[]{\label{sub7} \includegraphics[width=1.56cm]{figs/peppers_gray.pdf}}
    \subfigure[]{\label{sub8} \includegraphics[width=1.56cm]{figs/cameraman.pdf}}
    \subfigure[]{\label{sub9} \includegraphics[width=1.56cm]{figs/lena_gray_512.pdf}}
\subfigure[]{\label{sub10} \includegraphics[width=1.56cm]{figs/goldhill.pdf}} 
    \caption{Original images: (a) Mandril, (b) Peppers, (c) Cameraman, (d) Lena, (e) Goldhill. Watermarked images : (f) Watermarked Mandril, (g) Watermarked Peppers,(h) Watermarked Cameraman, (i) Watermarked Lena, (j) Watermarked Goldhill.}
    \label{stillimages}
\end{figure}

\begin{table}[!t]
\footnotesize
\centering
{\renewcommand{\arraystretch}{1.2}
\caption{Watermark imperceptibility measured in terms of PSNR (dB).}
\label{tab:imperceptibility1}
\begin{tabular}{|c||c|}
\hline
Images & \: Proposed Scheme  \: \tabularnewline
 \hline
Mandrill & \textbf{61.28} \tabularnewline
\hline
Lena & \textbf{61.97} \tabularnewline
\hline
Peppers & \textbf{65.97} \tabularnewline
\hline
\: Cameraman \: & \textbf{63.54} \tabularnewline
\hline

\: Goldhill \: & \textbf{66.37} \tabularnewline
\hline
\end{tabular}}
\\

\end{table}

\begin{table}[t]
\footnotesize
\centering
{\renewcommand{\arraystretch}{1.2}
\caption{Watermark imperceptibility measured in terms of PSNR (dB).}
\label{tab:imperceptibility2}
\begin{tabular}{|c||c||c|}
\hline
Images & \: Proposed Scheme  \: & \: Scheme in \cite{behloul2014blind} \:  \tabularnewline
 \hline
Mandrill & \textbf{58.80} &  56.262 \tabularnewline
\hline
Lena & \textbf{59.28} & 56,653 \tabularnewline
\hline
Peppers & \textbf{61.29} & 57.644 \tabularnewline
\hline
\: Cameraman \: & \textbf{60.00} & ---\tabularnewline
\hline

\: Goldhill \: & \textbf{61.60} & ---\tabularnewline
\hline
\end{tabular}}
~\\
\end{table}

Table \ref{tab:imperceptibility1} shows the imperceptibility results in PSNR using a $(19 \times 52)$ logo as watermark.
From Fig. 4 and Table \ref{tab:imperceptibility1}, it can be seen that the watermarked images preserve good visible quality and thus there is no visual distortion. Besides, all the obtained PSNR values are above  $61$ dB.\par
In Table \ref{tab:imperceptibility2}, is presented the comparison in terms of imperceptibility between the proposed scheme and the scheme in \cite{behloul2014blind}.
 The results show the superiority of our method.
 
\subsection{Robustness against attacks}
To evaluate the quality of the extracted watermark, we use the Normalized Correlation (NC), which is defined by : 
\begin{equation}
NC = \frac{ \sum_{i=1}^{M}\sum_{j=1}^{N} \begin{bmatrix}
W(i,j) & \times W'(i,j)
\end{bmatrix}^2}
{\begin{pmatrix}
\sqrt{\sum_{i=1}^{P} \sum_{j=1}^{Q} \left [ W(i,j) \right ]^2}  &  \sqrt{\sum_{i=1}^{P} \sum_{j=1}^{Q} \left [  W'(i,j)\right ]^2}
\end{pmatrix}}
\end{equation}
Where, $W$ and $W'$ are the original and the extracted watermarks, respectively. \\
Before applying  attacks, it can be observed that the watermark was extracted perfectly with a correlation NC=1.

To test the algorithm robustness, the watermarked images are exposed to several kinds of attacks: 1) noising attack : Gaussian Noise and Salt \& Pepper noise; 2) format-compression attack : JPEG compression; 3) image-processing attack : Low-Pass Gaussian Filtering and Histogram Equalization.
4) Combined attack : Histogram Equalization \& Gaussian Noise . \par
\begin{figure}
    \centering
    \subfigure[]{\label{sub1} \includegraphics[width=2.4cm]{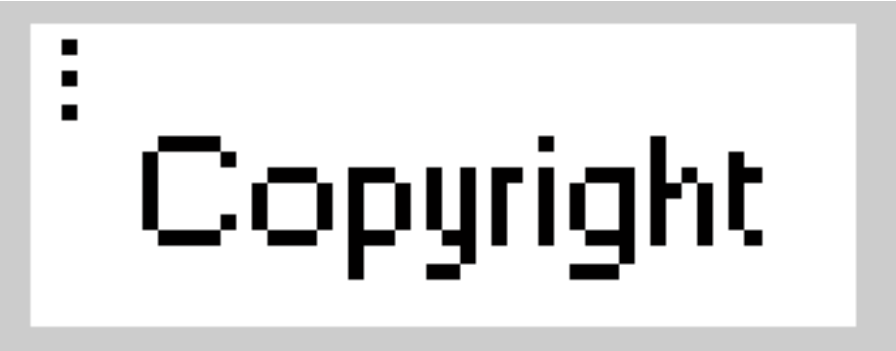}}
    \subfigure[]{\label{sub1} \includegraphics[width=2.4cm]{figs/marque_originale.pdf}}
    \subfigure[]{\label{sub1} \includegraphics[width=2.4cm]{figs/marque_originale.pdf}}
    \subfigure[]{\label{sub2} \includegraphics[width=2.4cm]{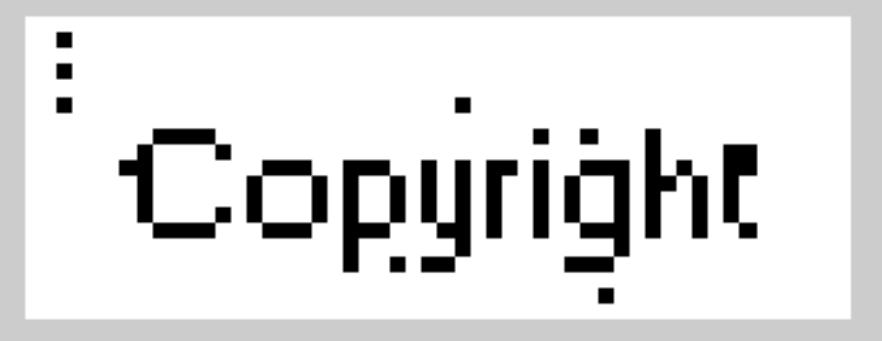}}
    \subfigure[]{\label{sub3} \includegraphics[width=2.4cm]{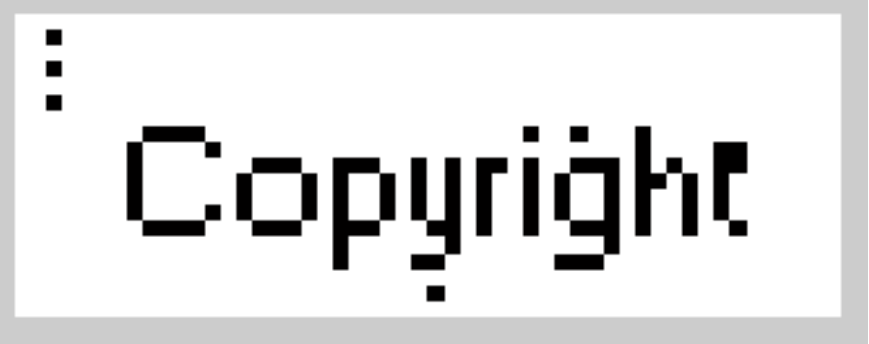}}
    \subfigure[]{\label{sub4} \includegraphics[width=2.4cm]{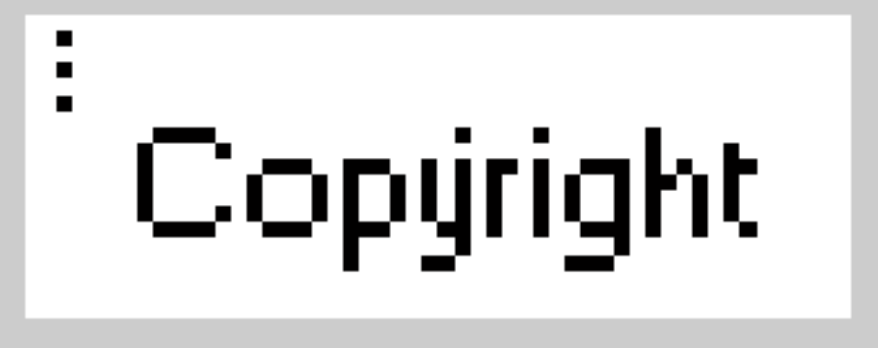}}
\subfigure[]{\label{sub6} \includegraphics[width=2.4cm]
{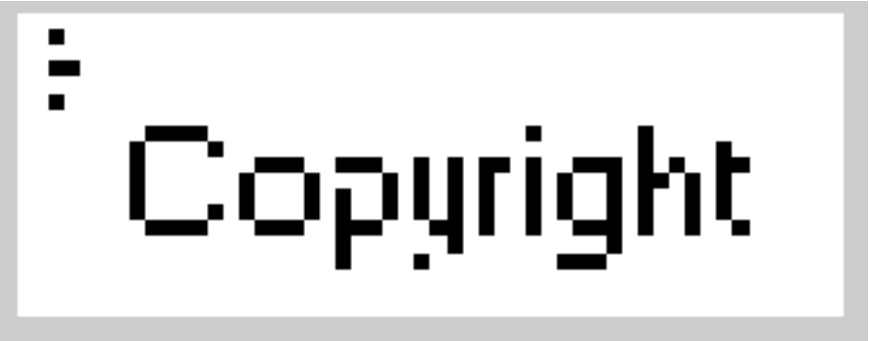}}
\subfigure[]{\label{sub5} \includegraphics[width=2.4cm]{figs/marque_originale.pdf}}

    \caption{Extracted watermarks after attacks : (a) Embedded Watermark,(b) Histogram Equalization, (c) Salt \& Pepper , (d) JPEG(Q=60) , (e) JPEG(Q=65)  , (f) JPEG (Q=70)  (g) JPEG (Q=75), (h) Gaussian Smoothing.}
    \label{stillimages}
\end{figure}

\begin{table*}[t]
\footnotesize
\centering
{\renewcommand{\arraystretch}{1.6}
{\setlength{\tabcolsep}{0.5cm} 
\caption{NC and PSNR values under various attacks.}
\label{tab:tableau-comparatif}
\begin{tabular}{|c|c||c|c||c|c|c|}
\hline
 & \multicolumn{2}{c|}  {Proposed scheme}  &  \multicolumn{2}{c|} {Scheme  \cite{mingzhi2013combined} } \tabularnewline
\cline{2-5}
  Attacks & \: PSNR (dB) \: & \: NC \:  & \: PSNR (dB) \: & \: NC \: \tabularnewline
\hline
\hline
No Attack  & Infinity & \textbf{1.0}  & Infinity & 1.0  \tabularnewline  
\hline
  Histogram Equalization & 18.87 & \textbf{1.0} & --- & ---  \tabularnewline
\hline
  Gaussian Noise ($\mu=0 $  $ ,\sigma=0.001$) & 34.59 & \textbf{1.0} & --- & 0.8895   \tabularnewline
\hline
  Salt \& Pepper ($\mu=0 $  $ ,\sigma=0.001$) & 34.60 & \textbf{1.0}  & --- & 0.8990  \tabularnewline
\hline
\: Histogram Equalization + Gaussian noise \: & 16.22 & \textbf{1.0} &--- & ---  \tabularnewline 
\hline
\end{tabular}}}
\end{table*}

\begin{table}[!t]
\footnotesize
\centering
{\renewcommand{\arraystretch}{1.2} 
\caption{PSNR and NC values after JPEG Compression attack : (QF=90), (QF=85), (QF=80), (QF=75), (QF=70), (QF=65) and (QF=60). }
\label{tab:jpegcompressionattack}
\begin{tabular}{|c||c||c|}
\hline
\: JPEG Compression \: & \:  PSNR(dB) \: & \: NC \: \tabularnewline
 \hline
(QF=90) & 35.63 & \textbf{1.0}
\tabularnewline
\hline
(QF=85) & 33.20 & \textbf{1.0} \tabularnewline
\hline
(QF=80)& 31.41 & \textbf{1.0}\tabularnewline
\hline
(QF=75) & 30.31 & \textbf{1.0}\tabularnewline
\hline
(QF=70) & 29.87 & \: \textbf{0.9840} \: \tabularnewline
\hline
(QF=65) & 29.14 & \: \textbf{0.9738} \: \tabularnewline
\hline
(QF=60) & 28.46 & \: \textbf{0.9343} \: \tabularnewline
\hline 
\end{tabular}}
\end{table}

Fig. 5. displays the extracted watermarks after several attacks ( Histogram Equalization , Salt \& Pepper noise, JPEG compression and Gaussian smoothing). We can see visually that although the watermarked images are exposed to these attacks, the watermarks are almost extracted perfectly.\\
We carried out the addition of Gaussian noise attack with zero mean ($\mu=0$) and several variance values ($\sigma$), then we applied a ​​' Salt \& Pepper' noise and we summarized the average NC values for the five test images. The results of Table \ref{tab:saltandpepperattack} show that our approach is more robust than the approach in \cite{behloul2014blind}.

\begin{table}[!t]
\footnotesize
\centering
{\renewcommand{\arraystretch}{1.2} 

\caption{ NC values after Salt \& Pepper attack. }
\label{tab:saltandpepperattack}
\begin{tabular}{|c||c||c|}
\hline
\: Salt \& Pepper \:  & \: Proposed Scheme \: & \: Scheme \cite{behloul2014blind} \: \tabularnewline
 \hline
 \:  $(\mu=0 $  $ ,\sigma=0.01)$ \:  & \:  \textbf{0.9981} \: & \: 0.9974 \:
\tabularnewline
\hline
\: $(\mu=0 $  $ ,\sigma=0.02)$ \:  & \: \textbf{0.9962} \: & \: 0.9960 \: \tabularnewline%
\hline
\: $(\mu=0 $  $ ,\sigma=0.04)$ \: & \: \textbf{0.9962} \: & \: 0.9881 \: \tabularnewline
\hline
\: $(\mu=0 $  $ ,\sigma=0.06)$ \:  & \: \textbf{0.9962} \: & \: 0.9834 \: \tabularnewline
\hline 
\end{tabular}}
\end{table}

In order to evaluate the robustness against JPEG compression, we compressed the watermarked images "Mandrill", "Peppers", "cameraman", "Lena" and "Goldhill" by different quality factors and we summarize the average NC values for the five test images. Afterwards, we compare our scheme to \cite{mingzhi2013combined} and \cite{behloul2014blind}.

\begin{table}[!t]
\footnotesize
\centering
{\renewcommand{\arraystretch}{1.2} 
\caption{PSNR and NC values after Low-Pass Gaussian Filtering attack : ( window size = $(3 \times 3)$, $(5 \times 5)$, $(7 \times7)$ and $(9 \times 9$)). }
\label{tab:lowpassgaussianfilteringattack}
\begin{tabular}{|c||c||c|}
\hline
\: Low-Pass Gaussian Filtering  \: & \: PSNR(dB) \: & \: NC \: \tabularnewline
($\sigma=0.5$) & & \tabularnewline
\hline
$(3 \times 3)$ & 45.80 & \textbf{1.0}\tabularnewline
\hline
$(5 \times 5)$ & 45.66 & \textbf{1.0} \tabularnewline
\hline
$(7 \times7)$ & 45.65 & \textbf{1.0}\tabularnewline
\hline
$(9 \times 9)$ & 45.65 & \textbf{1.0}\tabularnewline
\hline
\end{tabular}}
\end{table}

\begin{table}[!t]
\footnotesize
\centering
{\renewcommand{\arraystretch}{1.2} 
\caption{NC values after JPEG Compression attack : (QF=90), (QF=70) and (QF=50). }
\label{tab:jpegcompressionattackcomparaison1}
\begin{tabular}{|c||c||c|}
\hline
\: JPEG Compression \: & \:  Proposed Scheme \: & \: Scheme \cite{mingzhi2013combined} \: \tabularnewline
 \hline
(QF=90) & \textbf{1.0} & 0.9783
\tabularnewline
\hline
(QF=70) & \textbf{0.9981} & 0.9721 \tabularnewline
\hline
(QF=50) & \textbf{0.9384} & 0.9342 \tabularnewline%
\hline
\end{tabular}}
\end{table}

\begin{table}[!t]
\footnotesize
\centering
{\renewcommand{\arraystretch}{1.2} 
\caption{NC values after JPEG Compression attack : (QF=90), (QF=80), (QF=70), (QF=60) and (QF=50). }
\label{tab:jpegcompressionattackcomparaison2}
\begin{tabular}{|c||c||c|}
\hline
\: JPEG Compression \: & \:  Proposed Scheme  \: & \: Scheme \cite{behloul2014blind} \: \tabularnewline
 \hline
(QF=90) & \textbf{1.0} & 0.8710%
\tabularnewline%
\hline
(QF=80) & \textbf{0.9981} & 0.8680 \tabularnewline
\hline
(QF=70) & \textbf{0.9981} & 0.8710%
\tabularnewline%
\hline
(QF=60) & \textbf{0.9841} & 0.8720 \tabularnewline
\hline
(QF=50) & \textbf{0.9384} & 0.8630 \tabularnewline
\hline
\end{tabular}}
\end{table}

\begin{table}[!t]
\footnotesize
\centering
{\renewcommand{\arraystretch}{1.2} 
\caption{PSNR and NC values after  Gaussian Smoothing with several window sizes : ($3 \times$ 3), ($5 \times 5$), ($7 \times 7$), and ($9\times 9$). }
\label{tab:GaussianSmoothnig}
\begin{tabular}{|c||c||c|}
\hline
\: Gaussian Smoothing \: & \: PSNR(dB)\: & \: NC \:\tabularnewline
($\sigma=0.5$) & & \tabularnewline
 \hline
$(3 \times 3)$ & 48.41 & \textbf{1.0}\tabularnewline
\hline
$(5 \times 5)$ & 48.89 & \textbf{1.0} \tabularnewline
\hline
$(7 \times7)$ & 48.89 & \textbf{1.0}\tabularnewline
\hline
$(9 \times 9)$ & 40.45& \textbf{1.0}\tabularnewline
\hline
\end{tabular}}
\end{table}

Table \ref{tab:jpegcompressionattack} shows the  NC values, which are almost close to $1$, obtained after JPEG attack with their corresponding PSNR values in the case of "Mandrill". It can be seen from Table \ref{tab:jpegcompressionattackcomparaison1} and Table \ref{tab:jpegcompressionattackcomparaison2} that our method gives better results than  approaches in \cite{mingzhi2013combined} and \cite{behloul2014blind}.
\\ According to Table \ref{tab:tableau-comparatif}, it can be observed that our method is robust against noising attack and gives better results. Moreover, it shows high robustness against Histogram Equalization (NC=1).   
 
The watermarked image was filtered with a low-pass Gaussian filter using several window sizes ($(3 \times 3 )$, $(5 \times 5)$, $(7 \times 7)$ and $(9 \times 9)$).
From Table \ref{tab:lowpassgaussianfilteringattack}, it is clear that our approach is quite robust to low-pass Gaussian  filtering. The results show that the robustness  is still good even with larger sizes of the filters (Filter $9 \times 9$, NC=1). In addition, from Table \ref{tab:GaussianSmoothnig}, it can be seen that our method is very robust against Gaussian smoothing attack for several filter sizes ( $(3 \times 3 )$, $( 5 \times 5)$, $(7 \times 7)$ and $(9  \times  9)$).

 Finally, we have combined different attacks (Histogram Equalization and Gaussian noise ) to test the robustness of our method. Results presented in table \ref{tab:tableau-comparatif} are encouraging and  proved that this method is still robust.
\section{Conclusion and Future Work}
A blind  robust image watermarking scheme for Copyright protection has been presented in this paper. The watermark is embedded in the middle band DCT coefficients of the DFT magnitude of cover image using a secret key for increasing security. 
 Results obtained in terms of robustness and imperceptibility show that our scheme can improve the resistance to a wide range of attacks, e.g., Histogram Equalization, Gaussian noise, Salt \& Pepper noise, Low-Pass Gaussian filtering, Gaussian smoothing, JPEG Compression, etc. Our future work will be focused on improving the resistance to other kind of attacks such as scaling, rotation, etc.
    
    \section*{ACKNOWLEDGMENTS}  
 
 This work has been supported by the project CNRS-CNRST STIC 02/2014.
\bibliographystyle{IEEEtran}
\bibliography{IEEEabrv,biblio}
\end{document}